\documentclass{eas}
\usepackage{graphicx}
\begin{document}

\title{FS\,CMa type binaries}
\author{Anatoly S. Miroshnichenko}\address{Department of Physics and Astronomy,
University of North Carolina at Greensboro, P.O. 26170, Greensboro,
NC, 27402--6170}
\author{Sergey V. Zharikov}\address{Instituto de Astronom\'ia, Universidad Nacional Aut\'onoma
de Mexico,  Ensenada,  Baja California, Mexico, 22800}

\begin{abstract}
FS\,CMa type stars is a group of $\sim$70 objects formerly known as
unclassified stars with the B[e] phenomenon. Their very strong
emission-line spectra in combination with a nearly main-sequence
luminosity suggest the binary nature for them. They possess strong
IR excesses due to radiation of circumstellar dust that implies a
compact distribution probably in a circumbinary disk. Our long-term
spectroscopic monitoring revealed neutral metal lines, which always
include that of Li {\sc i} 6708 \AA, in the spectra of some FS\,CMa
objects indicating the presence of a cool star. We present a summary
of our results with a first overview of FS\,CMa type binaries and
review possible implications for the nature and evolutionary status
of the entire group.
\end{abstract}
\maketitle

\section{Introduction}
One of the first optical spectroscopic and near-IR photometric
surveys conducted almost 40 years ago showed that $\sim$10\% of
$\sim$700 hot emission-line stars exhibited forbidden lines and
strong IR excesses due to radiation of circumstellar dust (Allen \&
Swings \cite{as76}). This unusual group was called ``peculiar Be
stars'' or Bep stars to distinguish them from classical Be stars
which showed smaller IR excesses due to radiation from circumstellar
gas. Another name for the group, B[e] stars, proposed by Conti
(\cite{c76}) only reflects the presence of forbidden lines and not
of the IR excess, which is found in the majority of these objects.
After $\sim$20 years of studies, which were not intense due to
expected heterogeneity of the group members, Lamers \etal\
(\cite{l98}) concluded that the B[e] phenomenon is observed in four
classes of stars with known evolutionary status: pre-main-sequence
Herbig Ae/Be stars, symbiotic binaries, compact Planetary Nebulae,
and some supergiants (e.g., $\eta$ Car). At the same time, 30
original group members were declared unclassified objects with the
B[e] phenomenon.

The main problems with the unclassified objects were lack of
spectral lines from the B--type star atmosphere due to a strong line
emission and circumstellar continuum radiation as well as sparse
data that hampered definite classification. Their properties were
critically evaluated by Miroshnichenko (\cite{m07}), who summarized
the main group features as follows: 1) early--B to early--A type
optical continuum with strong emission lines of hydrogen, Fe {\sc
ii}, [O {\sc i}], and sometimes of [Fe {\sc ii}] and [O {\sc iii}]
(absorption lines from the hot star atmosphere may be present as
well, but they are frequently veiled by the circumstellar
continuum); 2) a large IR excess that peaks at 10--30 $\mu$m and
sharply decreases longward; 3) location outside of star-forming
regions; and 4) a secondary companion which can be a fainter and
cooler star or a degenerate object. Several possibilities for their
evolutionary status were rejected, and the group was renamed to
FS\,CMa type objects (after a prototype object with the B[e]
phenomenon, Swings \cite{s06}). In particular, the pre-main-sequence
status was rejected due to their location out of star-forming
regions and a steeper decrease of the mid- and far-IR flux with
wavelength than that of young stars. Also the FS\,CMa objects are
most likely not a kind of proto-Planetary Nebulae, which show much
smaller near-IR excesses.

Below we address the current view of the FS\,CMa group as binary
systems and show some recent results of our studies of individual
objects.

\section{Nature of the FS\,CMa group and its current size}

The FS\,CMa group initially contained 23 objects from Allen \&
Swings (\cite{as76}). Seven objects remaining from the 30
unclassified ones still has an uncertain status due to lack of data.
The group was enlarged by 10 objects found in the {\it IRAS}
database by cross-identification with catalogs of optical positions
(Miroshnichenko \etal\ \cite{m07a}). Another 20 candidates were
found by Miroshnichenko \etal\ (\cite{m11}) in the Hamburg survey of
emission-line stars (Kohoutek \& Wehmeyer \cite{kw99}), which was
cross-correlated with the NOMAD catalog (Zacharias \etal\
\cite{z04}). Currently $\sim$20 more candidates found in NOMAD using
several photometric criteria (e.g., Miroshnichenko \etal\
\cite{m07a}) are being observed spectroscopically to confirm the
presence of the B[e] phenomenon, and the search continues.

Miroshnichenko (\cite{m08}) found that the H$\alpha$ emission in
most FS\,CMa objects is over an order of magnitude stronger than
that of classical Be stars of the same spectral type. This requires
mass loss rates of \.M $\ge 10^{-7}$ M$_{\odot}$\,yr$^{-1}$
considering they are single stars. Theory predicts such rates only
for single supergiants with L $\ge 10^5$ L$_{\odot}$ (Vink \etal\
\cite{v01}). Therefore, it is natural to assume that FS\,CMa objects
are binary systems which experienced mass transfer due to a Roche
lobe overflow phase in at least one of the components. The mass
transfer should be non-conservative to explain a large amount of
circumstellar dust in the systems.

\section{Detected binaries among the FS\,CMa objects}

Although the binary scenario seems very likely, it has been uneasy
to confirm binarity of the group objects. The most obvious one is
MWC\,623 with a composite spectrum of a mid--B type primary and a
K--type secondary of a similar brightness (Zickgraf \cite{z01}).
Much weaker signs of a composite spectrum have been detected in 10
other objects. Orbital periods have been measured only in two of
them: CI\,Cam (the He {\sc ii} 4686 \AA\ line, 19.41 days, Barsukova
\etal\ \cite{b06}) and MWC\,728 (B5e+G8, metallic lines, 27.5 days,
Miroshnichenko \etal\ \cite{m15}). The other 8 objects with a cool
star lines, including that of Li {\sc i} 6708 \AA\ whose origin in
evolved stars is still under debate, are: V669\,Cep (Miroshnichenko
\etal\ \cite{m02}), FX\,Vel, AS\,174, IRAS\,00470+6429,
IRAS\,07080+0605, IRAS\,17449+2320, IRAS\,07377$-$2523
(Mi\-rosh\-nichenko \etal\ \cite{m07a}), and AS\,386 (Miroshnichenko
\etal\ \cite{m11}). Binarity of the three brightest FS\,CMa objects
(FS\,CMa, HD\,50138, and HD\,85567) was detected by
spectro-astrometry (Baines \etal\ \cite{Baines2006}), but neither
spectroscopy nor interferometry have confirmed that yet.

\begin{figure}
 \includegraphics[height=5.5cm,width=5.5cm]{Miroshnichenko_POE2015_fig1a.eps}
 \qquad
 \includegraphics[height=5.6cm,width=5.6cm]{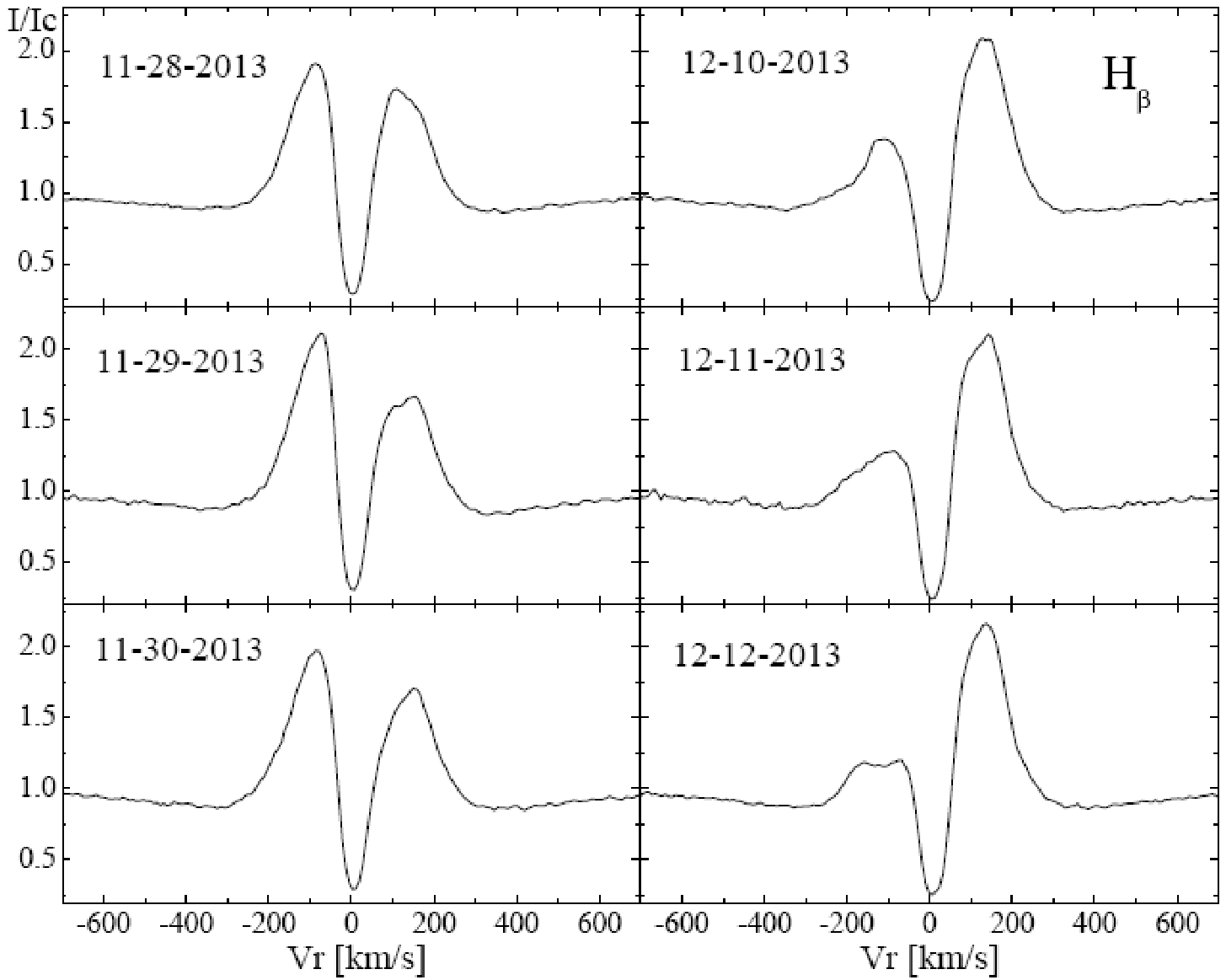}
 \caption{Fast variations of the H$\alpha$ line in the
 spectrum of MWC\,728 (left panel) and of the H$\beta$ line in the
 spectrum of HD\,50138 (right panel). Observations were obtained at
 the Observatorio Astronomico Nacional San Pedro Martir (Baja
 California, Mexico) and the Three College Observatory (North
 Carolina, USA) with spectral resolving powers of $R$ =
 10000--18000.}
\end{figure}

\section{Other features and problems}

Our recent frequent spectroscopic observations of some FS\,CMa
objects revealed significant variations of the Balmer line profiles
on a timescale of days which do not seem to be regular (see
Fig.\,1). This implies variable mass loss from the hot component,
although ongoing mass transfer due to a Roche lobe overflow has not
been confirmed yet. These irregular variations may veil regular ones
and hamper binarity confirmation. Nevertheless, the detected
composite spectra and the measured orbital periods are consistent
with non-conservative evolutionary models of binary systems (van
Rensbergen \etal\ \cite{vr08}). One of the future goals of the
FS\,CMa group investigation is to search for objects at different
stages of the binary evolution to identify when the dust formation
begins and evaluate the group contribution to the Galactic dust
production cycle.

A recent discovery of two FS\,CMa objects, which showed no signs of
a cool secondary in the near-IR region, in Galactic clusters (de la
Fuente \etal\ \cite{df15}) was interpreted as a consequence of
binary mergers. Although this idea seems possible, the secondary may
still be undetected due to its faintness with respect to the B--type
primary and veiling by the circumstellar continuum. In any case,
this result only enhances support for the binary nature of the
FS\,CMa group.

\section{Acknowledgments}

We acknowledge support from DGAPA/PAPIIT project IN100614. AM
acknowledges travel support from the University of North Carolina at
Greensboro.

\end{document}